\ifx\pdfoutput\undefined
   \documentclass[a4paper,12pt]{iopart}
\else
   \documentclass[a4paper,12pt,pdftex]{iopart} 
\fi
\usepackage{graphicx,iopams,fancyhdr}
\usepackage{setstack}
\usepackage{xy}
\usepackage{amsfonts}
\usepackage{url}
\newcommand{\leftshow}[2]{
 \includegraphics[width=#1]{#2}
}

\newcommand\centreshow[2]{
\begin{center}
\leftshow{#1}{#2}
\end{center}
}

\newcommand\centrefig[3]{
\begin{figure}
\centreshow {#1}{#2}
\caption{#3}
\end{figure}
}

\renewcommand\vec[1]{ {\mathbf #1} }

\newcommand\cross{\times}
\newcommand\grad{ {\bf \nabla } }

\newcounter{problemnumber}

\def\prl{\emph{Physical Review Letters}}

\def\jfm{\emph{J. Fluid Mechanics}}
\def\jpa{\emph{J. Physics A: Mathematical and General}}

\newcounter{theorem}
\newenvironment{theorem}[1]{\addtocounter{theorem}{1}\par
 \textbf{Theorem \thetheorem} \emph{ #1}
 }{\par\smallskip}

\newcounter{proof}

\newcommand\calug{C\u{a}lug\u{a}reanu~}

\newcommand\tvec{\widehat{\vec T}}

\newcommand\uhat{\widehat {\mathbf U}}
\newcommand\vhat{\widehat {\mathbf V}}

\newcommand\axis{\vec x}

\newcommand\axiscurve{\axis}
\newcommand\secondary{\vec y}

\newcommand\link{ {\mathcal L}}
\newcommand\lwig{ \widetilde{\mathcal L}}

\newcommand\twist{{\mathcal T}}
\newcommand\writhe{{\mathcal W}}

\newcommand\mutualsum[4]{\underset{#1\ne#3}{\sum_{#1=1}^#2
\sum_{#3=1}^#4}}
\begin{document}
\title{The evaluation of directionally writhing polymers.}
\author{C B Prior$^1$ and M A Berger$^2$}
\address{$^1$Department of Mathematics, University College London, University College London, Gower Street, London, WC1E 6BT, UK}
\ead{chriszprior@hotmail.co.uk}
\address{$^2$Mathematics, SECAM, University of Exeter, North Park Road, Exeter EX4 4QE, United Kingdom}
\ead{m.berger@exeter.ac.uk}
\begin{abstract}
We discuss the appropriate techniques  for modelling the geometry of open ended elastic polymer molecules.
The molecule is assumed to have fixed endpoints on a boundary surface.  In particular we discuss the concept of the winding number, a directional measure of the linking of two curves, which can be shown to be invariant to the set of continuous deformations vanishing at the polymer's end-point and which forbid it from passing through itself. This measure is shown to be the appropriate constraint required to evaluate the geometrical properties of a constrained DNA molecule. Using the net winding measure we define a model of an open ended constrained DNA  molecule which combines the necessary constraint of self-avoidance with being analytically tractable. This model builds upon the local models of Bouchiat and M\'ezard (2000). In particular, we present a new derivation of the \emph{polar writhe} expression, which detects both the local winding of the curve and non local winding between different sections of the curve.
We then show that this expression correctly tracks the net twisting of a DNA molecule subject to rotation at the endpoints, unlike other definitions used in the literature.
\end{abstract}

\section{Introduction}
The linking of a pair of space curves has been quantifiable since Gauss, who developed an expression for evaluating the integer degree to which the paths of two celestial bodies were linked (Epple \cite{epple}). In contemporary study this measure has been widely used to characterize the helical inter-linking of the two phosphate backbones of the DNA molecule (see Bates and Maxwell \cite{dnabook} for an overview of the role played by topology in the analysis of DNA). The geometrical study of DNA has been greatly assisted by a theorem of \calug (\calug \cite{calgal1}, \cite{calgal2}, White \cite{white} Fuller 1971 \cite{fuller0}) which separates the linking into components, representing the self linking of the helix's central axis (the \textit{writhe} $\writhe$) and the number of times the backbones wind about this axis (the \textit{twist} $\twist$). Denoting the linking as $\link$ \calug's theorem gives us
\begin{equation}
 \link =\twist +\writhe.
\end{equation}
We must state at this point that this theorem was defined for \textbf{closed} pairings (e.g., DNA Plasmids) and is not, in general, well defined when the constituent curves are open-ended (though in certain cases it can be shown to hold, van der Heijden et al \cite{vanderheijden2}). Writhing has a special place in the study of DNA as it is used to quantify the phenomenon of supercoiling, in which the DNA molecules axis entangles with itself such that long DNA molecules can occupy a small area in the structure of a cell (\Fref{dnasupercoil}). 

\centrefig{6cm}{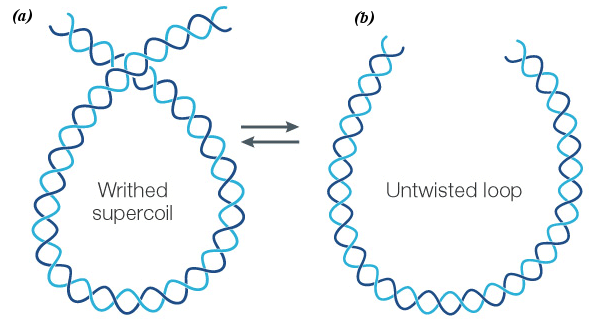}{\label{dnasupercoil} An example of supercoiling, of a section of DNA. The two figures are representations of a section of DNA molecule. In (a) the DNA's axis coils around itself to form a loop type structure. This loop is said to have writhing. This writhing is not present in (b). The two figures are interchangeable by an appropriate set of deformations applied to the axis. This figure is reprinted from Travers and Muskhelishvili \cite{supercoilpic}.}

\begin{figure}
\begin{centering}
\includegraphics[width=10cm]{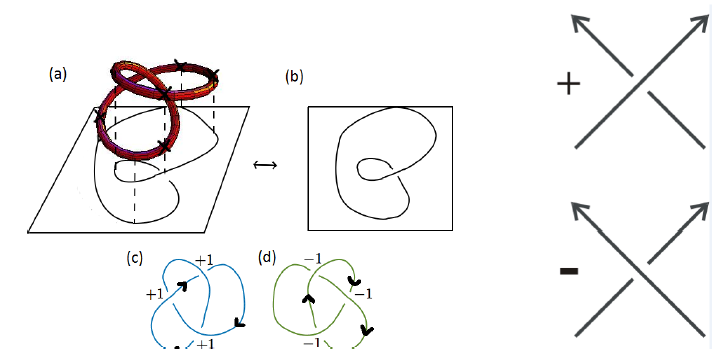}
 \caption{\label{writhecross} The diagram (a) represents the planar projection of a single space curve along a direction $\hat{b}$. Diagram (b) is the result of this projection as viewed along the direction of projection. Diagrams (c) and (d) represent example planar writhing calculations, using the crossing rules shown to the right.}
\end{centering}
\end{figure}

\centrefig{6cm}{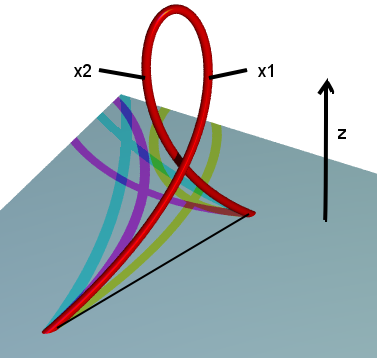}{\label{loop} A looped section of curve is split into two sections by its maximal point along $\hat{z}$. These sections are labelled $\axiscurve_1$ and $\axiscurve_2$. The two sections of curve which are separated by the local maximum in $z$, can be seen to wind around each other in the $x$-$y$ plane. Such contributions to the writhing geometry of the curve are non local and  are ignored by the Fuller writhing expressions.}

 If we consider a space curve $\axiscurve(t)$ (where $t$ is and arbitrary parametrization), which could represent the helical axis of a DNA molecule, its writhe $\writhe$ can be quantified as 
\begin{equation}
\label{writhe}
    \writhe \equiv \frac 1 {4 \pi} \oint_\axiscurve \oint_\axiscurve \frac{{\tvec_{\axiscurve}(t)} \cross {\tvec_\axiscurve(t')} \cdot (\axiscurve(t) - \axiscurve(t'))}{\vert\axiscurve(t)-\axiscurve(t')\vert^3} \rmd t\rmd t',
\end{equation}
where $\tvec_{\axiscurve}(s)$ is its the unit tangent vector of $\axiscurve$. Note that this expression is independent of parametrization. One can interpret \eref{writhe} as a directional average as follows. Consider a Cartesian direction $\hat{b}$ representing a viewing direction. We project the curve onto a plane perpendicular to this viewpoint as shown in  \Fref{writhecross}(a), with overlapping sections shown drawn above or below each other as demonstrated in the figure. If the curve is oriented we can apply a sign to each crossing as depicted in \Fref{writhecross}. The sum of these signs defines a measure of the planar writhing for a particular projection (see examples in \Fref{writhecross}(c, d)). The $\writhe$, as defined by \eref{writhe} can be shown to be the average over all directions of projection (see Pohl 1968 \cite{pohloriginal} or Aldinger \textit{et al} \cite{aldinger}).  As a result of this non-locality its use in analytic calculations can be limited (Bereton and Shah \cite{brereton1}). In 1978 Fuller \cite{fuller} introduced two single integral expressions for evaluating the writhing of closed space curves, which are defined in terms of the rotation of an orthonormal trihedron whose orientation is defined at all points by the local geometry of the curve $\axiscurve$. However, both expressions can only accurately evaluate the writhing of the axis $mod\,\,2$.  This is due to their ignorance of non-local (global) writhing (see \fref{loop}). That said, if certain necessary, but rather geometrically complex, conditions are met they can be used for both analytic calculation and efficient numerical evaluation (Starostin \cite{starostin}, van der Heijden \textit{ et al }\cite{vanderheijden2}). We do not discuss these conditions here as the aim is to provide a framework for evaluating the geometry of curve for which such restrictions are not necessary. 
\centrefig{8cm}{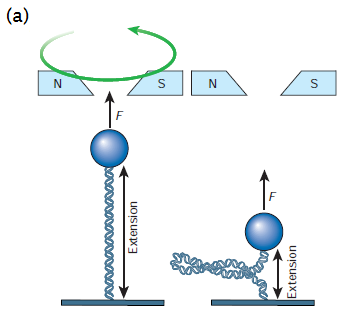}{\label{bustamante}A depiction of optical tweezer experiments. The molecules can be manipulated by optical tweezers. The magnets are used to manipulate the paramagnetic bead. The tow applied forces are a Torque force imparted by rotating the bead and s vertical force stretching the molecule. The right figure is a plectonemic structure which results form the application of a significant Torque. This figure is reprinted from Bustamante \textit{et al} \cite{bmexp2}.}
\centrefig{6cm}{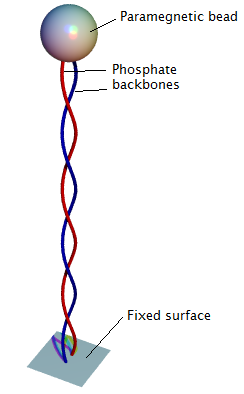}{\label{experiment}A depiction of typical set-up used to manipulate a DNA molecule in experiments. The molecule depicted as two phosphate backbones wound into a helical shape, typical of DNA molecules, is attached to a fixed surface and a paramagnetic bead. The bead  can be rotated (about a specific axis) by a magnetic field in order to apply a torque to the DNA molecule.}

The development of magnetic tweezer and optical trap techniques have allowed researchers to investigate the \textit{in vitro} response of DNA molecules to torsional stress (see Smith \textit{et al }\cite{bmexp1}, Bustamante \textit{et al }\cite{bmexp2} and Bustamante \textit {et al} for a review \cite{dnareview}). Of particular interest has been experiments in which an open ended section of a DNA molecule is held fixed at one end and manipulated at its second end by a paramagnetic bead (see \Fref{bustamante}). The molecule is subject to both a stretching force along this vertical direction and a torque perpendicular to the stretch force. We emphasize at this point that the forces controlling the molecule are either along or perpendicular to a specific Cartesian direction. Attempts to match the findings of these experiments have been the concern of a number of theoretical papers. These models focus on a polymer chain and require a partition function evaluation over the set of allowed geometrical configurations available to the molecule. In the limit of high force the molecules elastic behaviour is well understood using the \textit{worm-like chain model} (Bustamante \textit{et al }\cite{bmexp2}, Vologodskii \cite{volosolo}). Here the molecular axis only deviates slightly from pointing along $\hat{z}$ (the deviations are a result of random thermal fluctuations). There is currently more interest in the so called entropic elastic domain in which the set of allowed configurations can involve a notable degree of writhing (Vologodskii and Marko \cite{mandv}, Bouchiat and M\'ezard \cite{bandm}, Moroz and Nelson \cite{mandn}, Sinha \cite{sinha}). In particular Bouchiat and M\'ezard comprehensively detailed a model \footnote{Termed the Rod like chain (RLC) model  by the authors, but also referred to as the south avoiding worm like chain in Samuel et al \cite{indians}.}, which was shown to provide a good fit to the experimental data in this domain (similar results were also obtained by Moroz and Nelson \cite{mandn}). 

The aim in such statistical models is to identify all possible energy partitions (a continuous set in such macroscopic models, see Bouchiat and M\'ezard \cite{bandm}). In order to do this Bouchiat and M\'ezard devised an expression for the axial writhing of the molecule in terms of Euler angle rotations. One interpretation of this measure is an application of the local Fuller writhing expressions, though in a different form from that originally specified by Fuller (\cite{fuller}). It defines the local writhing of the molecule's axis about a specific direction (in their paper it is the vertical direction $\hat{z}$). In Cartesian form the expression (which we here denote $\writhe_z$) is the rate at which an orthogonal frame rotates about $\hat{z}$:
\begin{equation}
\writhe_z(\axiscurve)= \frac{1}{2\pi}\int_{0}^{M}\frac{\hat{z}\cdot(\tvec_1(t)\cross \dot{\tvec_t(t)})}{1+\hat{z}\cdot\tvec_t(t)}\rmd t.
\label{fuller2z}
\end{equation}
 A requirement, resulting from the ill definition of the Euler angles at $\theta=\pi$, is that the axis tangent curve is restricted from pointing along $-\hat{z}$ anywhere along is length. It appears to be the authors' claim that this restriction allows the \textit{local writhing} expression to replicate the behaviour of the better understood non-local $\writhe$ expression \eref{writhe}. In doing so the claim is made that this model of south-avoidance, that is the molecule's axis cannot point vertically downwards at any point along its length, will accurately capture the geometry of self avoidance (the appropriate physical obstruction to the systems global geometry). It is this particular issue which has led to a significant degree of debate as to the model's validity. Several papers, framed from a geometrical view point, have demonstrated that the local writhing expression does not accurately replicate the behaviour of the total writhing expression \eref{writhe} (Rossetto and Maggs \cite{rosmaglett} and in particular Neukirch and Starostin \cite{nands} and \cite{nands2}). On the other hand Samuel \textit{et al} \cite{indians} and \cite{indians2} take the view that the configurations, for which \eref{fuller2z} is in error, offer a negligible contribution to the full partition space, thus explaining the apparent success of Bouchiat and M\'ezard's model. We intend to show that this discussion is unnecessary as a more suitable writhe expression can be employed for open-ended, restricted polymers. 

This note contains the derivation the  path integral based model for defining and evaluating the appropriate non-local partition function of an open ended polymer, such as the DNA molecules studied in the micro-manipulation experiments we have discussed. With regards to the experiments, a key factor in the modelling of the molecule is the directional nature of the applied forces. The forces applied to the field are transmitted through control of the paramagnetic bead (\fref{experiment}) to which the upper end of the molecule is attached. This bead is placed in between the north and south poles of a magnet. By altering the polarity of this magnet the bead can be rotated by applying a torque force to the molecule. This force is always perpendicular to a chosen direction. In this note we shall chose this direction to be vertical ($\hat{z}$) following the convention used by Fain et al \cite{fain}  and Bouchiat and M\'ezard \cite{bandm}. That is to say the bead rotates in the $x$-$y$ plane. Secondly a force $F$ can be applied to the bead stretching it along $\hat{z}$. 

The paper will take the following structure. In \sref{energysec} we discuss the energy expression of an elastic ribbon representing the polymer molecule. The expression is derived by considering the rates of rotation of an orthonormal frame attached to a curve representing the polymer's axis. In particular the twisting expression differs from that in the literature (\cite{fain},\cite{bandm},\cite{mandn}). However, it is shown that despite this difference the energy expression for the ribbon is the same as that used in the cited literature.  

In \sref{lkinvariant} we introduce the net winding expression. This is a measure of the directional linking of a pair of curves. A discussion relating this expression to the confined polymer experiments we have discussed in the introduction demonstrates that the net winding is the correct geometrical constraint, which should be used to restrict the allowed configurations of the ribbon. 

In \sref{polarmonopole} we detail the correct writhing expression for confined polymers. This measure was previously introduced using the net winding measure and is termed the polar writhe (Berger and Prior \cite{main}), however, it is derived here in a different manner. We use the correct form of a magnetic monopole field in order to derive the polar writhe. This derivation supplies an interpretation for the link between the quantum mechanical problem of a symmetric top in a magnetic field and a constrained polymer molecule, subject to torsional and stretching forces. This link was first suggested by Bouchiat and M\'ezard without a geometrical justification which we present here. Further to this we use two example curve studies, a helix and a deformed spacecurve whose end points are fixed, in order to demonstrate that the polar writhe measure has the required geometrical properties for constrained polymers.

In \sref{partition} we derive a partition function, using the net winding measure, which satisfies the non local self-crossing restriction, along with the possibility of being evaluated analytically. This is achieved by splitting the net winding measure into local and non local components. The local contribution is decomposed into directional twisting and winding contributions in a manner similar to that of Bouchiat and M\'ezard (though with differing twist and writhe expressions). As the energy expression is defined in terms of local geometry we use the non local winding contribution as a constraint which enforces the non locality of the model.   

\section{Defining the energy of the polymer}\label{energysec}
\subsection{Geometrical background}
In what follows a space curve $\axiscurve(t)$ shall represent a three-dimensional vector $(\axiscurve_x,\axiscurve_y,\axiscurve_z)$, depending continuously on an arbitrary parameter $t$, for $t\in[a,b]$.  The Euler angles in this note will  be the set ($\phi\,$,$\,\theta$,$\,\psi$) with $\phi$ a rotation in the $x$-$y$ plane, $\theta$ a rotation in the $y$-$z$ plane for $\hat{z}$  and $\psi$ a rotation in the $x$-$y$ plane, following the rotation $\theta$. In addition to the Cartesian system and the Euler angle set, defined in \sref{symtop}, we will use the spherical polar coordinate system $(r,\theta,\phi)$, with theta the polar angle and $\phi$ the azimuthal angle. It shall be made clear when we are discussing spherical polar or Euler angle parametrisations. All such curves will be considered \textit{smooth}, where smoothness implies $\axiscurve$ is at least $\mathcal{C}^3$ differentiable for all $t$. Differentials are denoted with a prime (ire $x'(t) =  \rmd x/\rmd t$).

\subsection{Ribbons}
\centrefig{15cm}{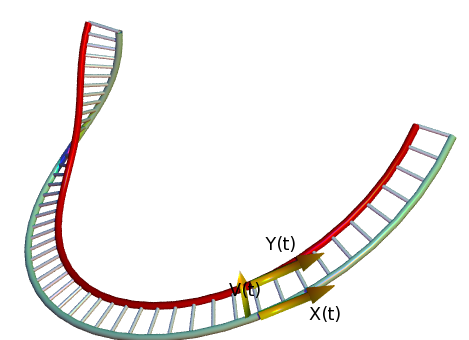}{\label{ribbonfig}A typical ribbon construct with $\vec{x}(t)$ representing the ribbon's axis. The vector $\vec{v}(t)$ generates the curve $\secondary(t)$ as defined by the equation $\secondary(t) = \axiscurve(t) + \epsilon\vec{v}(t)$.}

The ribbon is a mathematical construction used to represent polymers such a DNA molecules  (Fuller \cite{fuller}). Consider a space curve $\axiscurve(t)$ and a second curve $\secondary(t)$, also parametrized by $t$, such that $\secondary(t) = \axiscurve(t) + \epsilon\vec{u}(t)$, where $\vec{u}(t)$ is a vector normal to $\tvec(t)$ for $t\in [a,b]$. This will naturally wrap itself around $\axiscurve$ as shown in Figure \ref{ribbonfig}. If $\epsilon$ is sufficiently small (usually $\ll 1$) we can assume that $\secondary$ is disjoint from $\axiscurve$, that is $\axiscurve$ and $\secondary$ never cross (Hirsch \cite{hirsch}). Such a construction will be denoted $R(\axiscurve,\secondary)$. Closed ribbon's require $\secondary$ to be closed over the same period as $\axiscurve$.

\subsection{The enegry components}

In order to create our partition function we must define the energy required for the polymer to assume each possible configuration. In Bouchiat and M\'ezard the elastic energy of the polymer molecule was define using a set of rotations (parametrized using the Euler angles) of an orthonormal frame $\left( \tvec,\uhat,\vhat \right)$. The rotations were relative to a fixed rectilinear reference frame pointing vertically (along $\hat{z}$). Let $T_z =\tvec \cdot \hat z$  be the vertical component of the tangent vector, and $\theta = \cos^{-1} T_z$. Also let $\sigma$ give the sign of $T_z$, i.e. 
\begin{equation}
\label{sigmarules}
\sigma_i = \left\{
\begin{array}{c}
1,\,\,\,\, \mbox{ if }  0 \leq  \theta \leq \pi/2,\\
-1 \,\,\,\, \mbox{ if } \pi/2 < \theta \leq \pi.\\
\end{array}
\right.
\end{equation}
We will multiply our rotations by $\sigma$. A physical justification is given in Appendix A.
We shall see that, whilst this does affect the expression for the net twisting of the molecule, it does not affect the energy expression using in Fain \textit{et al} \cite{fain} or Bouchiat and M\'ezard \cite{bandm}.  However, we shall see later in \sref{partition} that the difference in twisting does have an effect when we apply the necessary geometrical bounds on the set of configurations accessible to the molecule.

 We define the angular velocity ($\vec{\Omega}$) of our orthonormal frame as
\begin{equation}
 \vec{\Omega} = \omega_1\tvec + \omega_2\uhat + \omega_3\vhat, 
\end{equation}
where the $\omega$ values are scalars representing the rates of rotation about that particular direction. For the sake of brevity we do not detail the derivations. Bouchiat and M\'ezard have already detailed the procedure for deriving the three angular velocity rates. We have repeated this procedure whilst applying the rule set defined in \eref{sigmarules} and find the following expressions 
\begin{eqnarray}
\omega_1
\label{omega1polar}&=& \frac{\rmd{\psi}}{\rmd{z}} + \vert\cos{\theta}\vert\,\frac{\rmd{\phi}}{\rmd{z}}.\\
\omega_2 &=& \sigma_i\frac{\rmd\theta}{\rmd{z}}.\\
\omega_3 &=& \sigma_i\sin{\theta}\frac{\rmd\phi}{\rmd{z}}.
\end{eqnarray}

\subsection{The energy expression}\label{energydensity}
In general the energy of a particular configuration of an elastic polymer, modelled as a ribbon, has three components. The first component results from the bending. Following the work of Fain \textit{et al} \cite{fain} and  Bouchiat and M\'ezard \cite{bandm} this can be quantified by the two angular velocity components $\omega_2$ and $\omega_3$ which represent the rate of rotation of the tangent vector about $\uhat$ and $\vhat$, leading to the following density,
\begin{equation}
E'_{bend} = \frac{A}{2}(\omega_2^2 + \omega_3^2)= \frac{A}{2}\left(\left(\frac{\rmd{\phi}}{\rmd{z}}\right)^2\sin^2{\theta} + \left(\frac{\rmd{\theta}}{\rmd{z}}\right)^2\right),
\end{equation}
where $A$ is the bending rigidity constant. The second contribution comes from the twisting of the polymer's edges about its central axis. This can be defined by the density
\begin{eqnarray}
E'_{twist} &=& \frac{C}{2}\omega_1^2 = \frac{C}{2}\left( \sigma_i\frac{\rmd{\psi}}{\rmd z} +\vert\cos{\theta}\vert\,\frac{\rmd{\phi}}{\rmd{z}}\right)^2,\\
&=& \frac{C}{2}\left(\frac{\rmd{\psi}}{\rmd z} +\cos{\theta}\,\frac{\rmd{\phi}}{\rmd{z}}\right)^2.
\end{eqnarray}
Here $C$ is the twist rigidity constant. Finally to complete our expression we allow for the molecule to be stretched. In this experiment the molecule is stretched along the $z$ axis. A uniform stretching force ($\vec{F} = F\cdot\hat{z}$) is applied at the free end of the molecule. This is a potential and acts at all points along the molecule based on their orientation. If the molecule is moving upwards in $\hat{z}$ then the force will have a negative value. If it is moving downwards it will have a positive value, giving
\begin{equation}
E'_{stretch} = - \frac{F\cos{\theta}}{k_bT}.
\end{equation}

\section{An invariant linking expression -- The net winding number}\label{lkinvariant}

\centrefig{8cm}{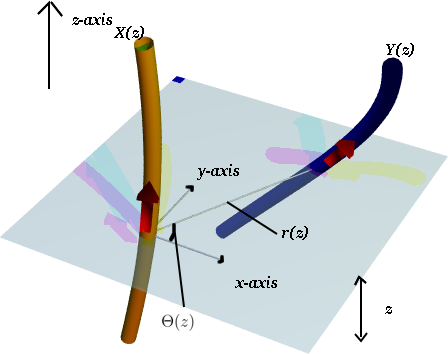}{\label{polarlink} Two sections of curves $\axiscurve$ and $\secondary$ occupying the same $z$ range. Their linking, along $\hat{z}$, can be defined in terms of the vector $\vec{r}(z)$ joining two sections of curve $\axiscurve(z)$ and $\secondary(z)$ in the $x-y$ plane. Also depicted is the angle $\Theta(z)$ which $\vec{r}$ makes with the $x$-axis. $\Theta$ is a combination of the two Euler angles $\phi$ and $\psi$.}

Having discussed the energy expression for our cylindrically symmetric elastic polymer we must now consider the geometrical restrictions placed on the molecule. It is at this juncture that our work begins to differ significantly from other treatments of the problem (Fain \textit{et al} \cite{fain}, Bouchiat and M\'ezard \cite{bandm}, Moroz and Nelson \cite{mandn}, Rossetto and Maggs \cite{rossetto}, Samuel \textit{et al} \cite{indians}). For this discussion we shall assume the  torque and stretching forces take on fixed values (though we assume they are arbitrarily chosen, such that this discussion is generalized). In this scenario we would like a measure of the polymers geometry, which remains fixed for fixed torque and stretching values. In Berger and Prior \cite{main} just such an invariant was defined, which we refer to as the net winding $\lwig$. It represents the extent to which the axis of the polymer and a second curve, (possibly representing one of the protein strands of a DNA molecule), wind about each other perpendicular to $\hat{z}$ (see \fref{polarlink}). It is of note that this expression is not in general equal to the Gauss linking number ($\link$), which is a well known topological invariant for \textbf{closed} polymer ribbons. Authors have suggested artificial extensions of the polymer-ribbon which allow the use of the Gauss linking invariant (Rossetto and Maggs \cite{rossetto}, Samuel \textit{et al} \cite{indians}), we show in this section that this extension is not necessary, as the appropriate invariant can be described without the need for an artificial extension. This measure will be split into two distinct contributions, the local winding $\lwig_l$ and the non-local winding $\lwig_{nl}$.

\subsection{The local winding}
Consider a section of the polymer ribbon $R(\axiscurve,\secondary)$, parametrized by $z$, for which both sections are travelling upwards in in $z$ ($\axiscurve'(z)> 0$ and $\secondary'(z) > 0$), over a range $z\in[z_{min},z_{max}]$.  We define a vector $\vec{r}$, originating at $\axiscurve$ which will lie in the $x$-$y$ plane such that its tip lies on $\secondary(z)$ for all $z\in[z_{min},z_{max}]$,
\begin{equation}
\vec{r}(z) = \secondary(z) -\axiscurve(z),
\end{equation}
(see \Fref{polarlink}). We wish to measure the rate of rotation of this vector in the $x$-$y$ plane, which can be used to represent the extent to which the two curves wind about each other, perpendicular to $\hat{z}$, over this range. In order to do so we must evaluate the rate of rotation of the vector as we increase $z$, in  Euler angle rotations this would be,
\begin{equation}
\label{eulerlocal}
\left(\frac{\rmd\psi}{\rmd{z}} + \frac{\rmd{\phi}}{\rmd z}\right)\rmd{z}.
\end{equation}
By integrating over $z$  we obtain the net winding of the ribbon $R(\axiscurve,\secondary)$ perpendicular to $\hat{z}$.
This result was obtained by Bouchiat and M\'ezard \cite{bandm} in a different fashion. The same expression applies as for sections of the polymer which are travelling downwards in $\hat{z}$  ($\axiscurve'(z)> 0$ and $\secondary'(z) > 0$). We have termed this expression the local winding $\lwig_{l}$ as all contributions arise form the winding of curve sections $\axiscurve_i$ and $\secondary_i$ belonging to the same section of the ribbon. 

\subsection{The non-local winding}
\centrefig{10cm}{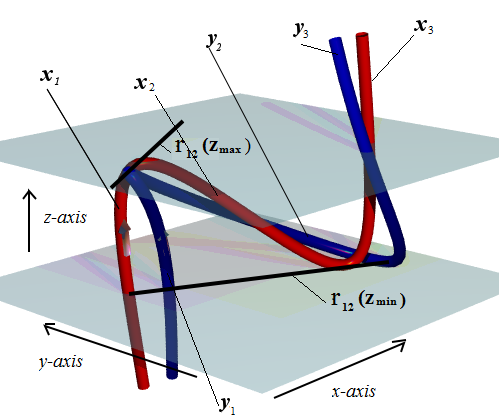}{\label{turnribbon}A depiction of a section of ribbon 
$R(\axiscurve,\secondary)$ which has both local winding and non local winding. Both curves are split into three sections ($\axiscurve_1$, $\axiscurve_2$, $\axiscurve_3$, $\secondary_1$, $\secondary_2$ and $\secondary_3$) by their turning points along the $z$-axis. Shown on the diagram are two planes $z=z_{min}$ and $z=z_{max}$ which bound the range of $z$ values shared by $\axiscurve_1$ and $\secondary_2$. Also depicted are the directions of $\vec{r}_{12}$ at $z=z_{min}$ and $z=z_{max}$.  we used these vectors to evaluate the non-local winding of the ribbon.}
In order to render this measure a topological invariant (to the set of deformations  vanishing at its endpoints, see \sref{invariance}) we must consider a further set of contributions due to the winding of the polymer between distinct sections of the polymer. Consider a polymer configuration which has two turning points in the $\hat{z}$ direction (\Fref{turnribbon}). There will be additional linkage of the two curves for the range of $z$ values which share a mutual $z$ range as marked on \Fref{turnribbon}. Consider for example the sections  marked $\axiscurve_1$ and $\secondary_2$.  In order to define the net winding of this pair we use a vector $\vec{r}_{12}$ (see \Fref{turnribbon}) which is defined as
\begin{equation}
\vec{r}_{12}(z) = \secondary_2(z) - \axiscurve_1(z).
\end{equation}
By defining the infinitesimal rate of rotation as described above we can recover the net winding between these two sections by integrating over their mutual $z$ range $z\in[z_{12}^{min},z_{12}^{max}]$,
\begin{equation}
\lwig_{12}(\vec{R},z_{12}^{min},z_{12}^{max}) = \frac{1}{2\pi}\int_{z_{12}^{min}}^{z_{12}^{max}}-\left(\frac{\rmd\psi_{12}}{\rmd z} + \frac{\rmd{\phi_{12}}}{\rmd{z}}\right)\rmd{z},
\end{equation}
where the labelling $\psi_{12}$ and $\phi_{12}$ indicate that they represent the required Euler rotations necessary for the top-vector combination to trace out the polymer section $R(\axiscurve_1,\secondary_2)$. The minus sign occurs as the two sections are travelling upwards and downwards along $\hat{z}$ respectively (see \Fref{turnribbon}). We can generalize this construction to cover all geometries which are possible for a polymer bound at two specific points, which are distinct (this is exactly the case for the DNA molecule discussed above). For a polymer constructed from curves $\axiscurve$ and $\secondary$ which have $n-1$ and $m-1$ turning points along $\hat{z}$ we consider the set of vectors $\vec{r}_{ij}$ attached to the top centred on section $\axiscurve_i$ which points to section $\secondary_j$. The total net winding of  a polymer $R(\axiscurve,\secondary)$ , about $\hat{z}$ between two planes $z_{min}$  and $z_{max}$, is then the sum of all contributions between sections $\axiscurve_i$ and $\secondary_j$,
\begin{eqnarray}
\lwig(\vec{R},z_{min},z_{max}) &=& \sum_{i=1}^n\sum_{j=1}^m\lwig_{ij},\\
&=&  \sum_{i=1}^n\sum_{j=1}^m\frac{\sigma_i\sigma_j}
{2\pi}\int_{z_{ij}^{min}}^{z_{ij}^{max}}\left(\frac{\rmd\psi_{ij}}{\rmd z} + \frac{\rmd{\phi_{ij}}}{\rmd{z}}\right)\rmd{z}.\label{netwinding}
\end{eqnarray}
Here the local contributions will be those for which $i=j$, which we have labelled $\lwig_l$. We call the contributions for which $i\neq j$ the \textit{non-local} contributions, we shall label these $\lwig_{nl}$. A key factor in our final model will be the distinction between the local and non-local contributions. 

\subsection{Topological invariance and directionality}\label{invariance}

\centrefig{6cm}{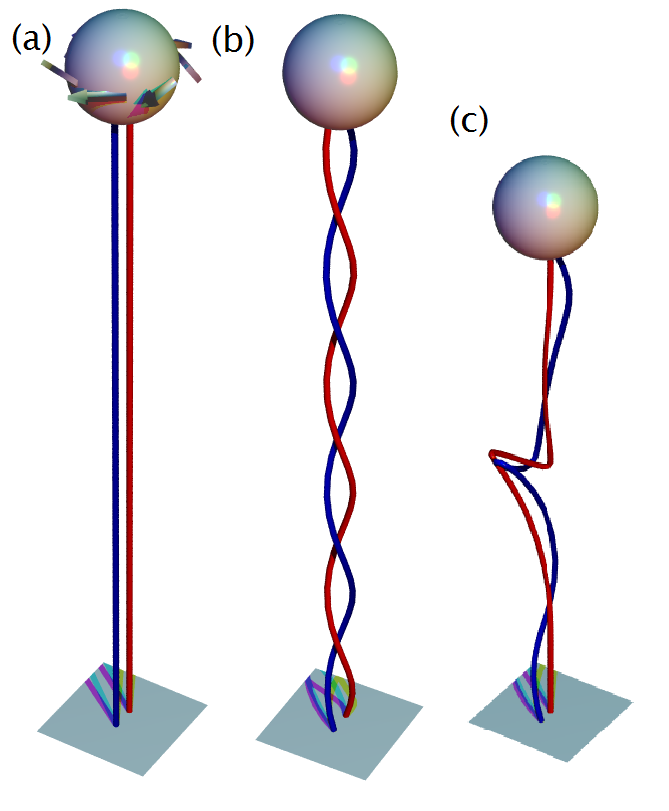}{\label{windingtypes}Depicted are three possible configurations of the constrained polymer (representing a DNA molecule for example). The configuration depicted in (a) is a rectilinear configuration which would only be present in the case in which there have been no turns of the bead and not linking of the polymer strands before attachment. The configuration depicted in (b) is a braided configuration which would result from twisting the bead as shown in (a). The number of windings of the two strands of the molecule about each other would be proportional to the number of turns applied to the bead (assuming we started with a rectilinear configuration). The molecule's axis is a straight line for (b) so it has no writhing. The figure in (c) is a configuration in which the number of turns applied to the bead is the same as (b). It has both non-local and local windings. In (c) the axis itself has a non-trivial geometry. Both configurations can be shown to have the same net winding.}

We can show that the net winding has a restricted sense of topological invariance, which is sufficient for open-ended ribbons whose endpoints are constrained. Before we discuss this theorem we introduce the concept of the base state. Fain \textit{et al} \cite{fain} show that the minimum energy state for such constrained polymers are braided vertical lines such as that marked (b) \Fref{windingtypes}. We shall call such states the base state. We assume that the curve can be arbitrarily deformed within some region of space, where the endpoints attach to the boundary of the region. What we shall show in the following theorem is that writhed configurations such as (c) (\Fref{windingtypes})  which result from deforming   the base state (b), subject to its end orientations being fixed, will have the same net winding.

\newcommand\zmin{z_{\min}}
\newcommand\zmax{z_{\max}}
In Berger and Prior the following theorem was proven. We restate it here such that we can use it in the following discussion to justify the use of the net winding as the correct topological restriction for constrained polymer molecules.
\begin{theorem}{\label{t1} 
Consider an continuous open ended ribbon $R(\axiscurve,\secondary)$ (both $\axiscurve$ and $\secondary$ are assumed to be continuous). We shall assume this ribbon is of at least $\mathcal{C}^3$ differentiability. Let $\zmin$ be the height of the lower endpoint and $\zmax$ the height of the upper endpoint. We restrict the motion of the ribbon to $\zmin< z < \zmax$. We define the set of restricted ambient isotopies, which vanish outside $\zmin< z < \zmax$, as the \textbf{end-restricted ambient isotopies}. Then the net winding, defined by equation \eref{netwinding} is invariant to all end-restricted ambient isotopies.}
\end{theorem}

For proof see Berger and Prior \cite{main}, the  pertinent theorem in that paper is  theorem 2, which in turn relies on theorem 1 (theorem 1 of \cite{main}, not the above) . Briefly the net winding of a ribbon of infinite length, is shown to be equal to the Gauss linking number (theorem 1 of \cite{main}). Consider a section of this curve (the section of interest) to which the deformations are constrained, and at whose end points the motions vanish. The net winding measure can be shown to be equal to the linking of the infinite assembly minus the net winding of the remaining ribbon sections (i.e. everything but the region of interest) which must be invariant as all motions vanish outside the region of interest. Thus the net winding of the section of interest is also invariant. 
 
It can be seen that, in terms of the systems physics, the net winding is the appropriate measure for evaluating the linking of the constrained DNA molecule. Consider a rectilinear polymer configuration as depicted in \Fref{windingtypes}(a). We then apply a Torque in the $x$-$y$ plane as in the experiments  (again see \Fref{windingtypes}(a)). This torque will cause the polymer to form the base-state braided structures (see \fref{windingtypes}(b)) with the number of turns of the braided structure corresponding to the the number of turns applied to the bead, this would naturally be measured by the net winding \eref{netwinding}. It is possible, given a significant number of turns of the bead, that the polymer structure can buckle to form structures such as \Fref{windingtypes}(c). These structures have significant writhing (we shall detail the appropriate expression for evaluating this writhing in \sref{polarmonopole}). What should be noted is that we have demonstrated in theorem 1 that the net winding will stay fixed, assuming the end points are prevented form rotating; the winding is merely distributed in a different manner.  It is for this reason that we use the net winding as means of con straining geometrical configurations available to the system, for a given number of turns of the paramagnetic bead. This suitability derives entirely form the directionally specific nature of both of the forces applied in the experiments. We must reiterate that the net winding is \textbf{not}, in general, equal to better known Gauss linking expression $\link$ which is defined as an average over all viewpoints and it  is physically appropriate, in this case, to use the net winding rather than the $\link$. This fact was already recognized by Bouchiat and M\'ezard who, derived their own directional winding expression. We have simply defined the appropriate form of the required directional invariant. 

 \section{Detecting the writhe using magnetic fields}\label{polarmonopole}
 
In this section we define the appropriate form of directional writhing of an open ended polymer's axis. In \sref{partition} this expression will be used in order to define an analytically implementable partition function which incorporates the self avoidance condition. This writhing expression, termed the polar writhe, has already been specified by Berger and Prior \cite{main}, using the net winding expression as a basis for construction. The paper was primarily written from a purely geometrical perspective. We choose to introduce this expression here in a different fashion, using the language of magnetic fields. Bouchiat and Me\'zard derived a version of the directional writhing expression which was shown to be equivalent to a line integral over the vector potential of a field generated by a thin magnetic solenoid which extends from infinity to some endpoint $\vec x$ (in effect this is the magnetic field of a Dirac monopole placed at position $\vec x$). They used this fact to develop a Hamiltonian expression based on a quantum symmetric top subjected to a magnetic field, though there was no real suggested physical interpretation for this link. In what follows we shall define the appropriate form of the directional writhing, utilizing the correspondence between electromagnetism and the geometrical language of fiber bundles first demonstrated by Wu and Yang \cite{wuyang}. In doing so we will assign a physical interpretation to this equivalence, with regards to the DNA-polymer experiments we are discussing.

We shall construct our writhing expression using a tethered magnetic monopole  (of charge $q=1$) and its vector potential. As magnetic monopoles do not possess globally defined vector potentials (and, as far as we know, do not exist in nature) we imagine a thin bundle of magnetic flux $\Phi = q/(4\pi)$ extending from $z = -\infty$ to the point $\vec x$, confined to a thin vertical tube. At $\vec x$ the flux escapes from the tube and spreads equally in all directions, mimicking a magnetic charge. Our set up will be equivalent to the rotated symmetric top description described in Appendix A (\sref{symtop}), with the charge replacing the top and the vector potential the attached directing vector. We consider a curve $\axiscurve$ which represents the axis of our polymer. We shall first consider a section of polymer moving only  upwards along $\hat{z}$. We start the magnetic monopole at the base of the polymers axis (minimum $z$ value) and attach the following vector potential 
\begin{equation}
\label{northvectorpot}
\vec{A}^N=\frac{\hat{z}\cross\vec{a}}{\vert \vec{a}\vert\,(\vert\vec{a}\vert+\hat{z}\cdot \vec{a})}.
\end{equation}
Here $\vec{a}$ represents a vector ($\axiscurve^*- \axiscurve$), with $\axiscurve$ being a point belonging to the field line and $\axiscurve'$ any other point between the planes. This vector potential represents one possible vector potential used to describe the Dirac magnetic monopole (Dirac 1931 \cite{dirac}). This potential is well defined for all $\axiscurve^*$ except points vertically below $\axiscurve$. For our purpose we shall demand that $\vec{a}$ is of unit length, further, we shall demand that for each $\axiscurve$ the point $\axiscurve^*$ will be such that $\vec{a}$ points along the direction of the unit tangent $\tvec$ giving
\begin{equation}
\vec{A}^N= \frac{\hat{z}\cross\tvec}{1+\hat{z}\cdot \tvec}
\end{equation}
We now consider the potential at a single point $\axiscurve$ and consider an infinitesimal change in the geometry of the field line $\rmd\tvec$. This will mean the field defined by the magnetic monopole will deform. We take the dot product of the potential and this infinitesimal change to define the rate of rotation of this potential 
\begin{equation}
\label{fieldint} 
\vec{A}^N\cdot\rmd{\tvec} =\frac{\hat{z}\cross\tvec}{(1+\hat{z}\cdot \tvec)}\cdot \rmd{\tvec}.
\end{equation}
A key geometrical property of the potential $\vec{A}^N$ is; whichever direction the field line is pointing the potential will always be projected into the $x$-$y$ plane. Thus the winding is defined in the $x$-$y$ plane, as we would require. As the field line $\axiscurve$ always has a positive gradient along $\hat{z}$ we can state that for each $z$ value the potential vector will be unique, that is none of the potential vectors will intersect. Using $\rmd{\tvec} = \tvec(z)'\rmd{z}$ we can rewrite \eref{fieldint} as an infinitesimal in terms of $\hat{z}$
\begin{equation}
\vec{A}^N\cdot\rmd{\tvec}  = \frac{\hat{z}\cdot(\tvec\cross\tvec')}{(1+\hat{z}\cdot \tvec)}\rmd{z}
\end{equation}
In spherical coordinates this gives
\begin{equation}
\label{northspherical}
\vec{A}^N\cdot\rmd\tvec = \left(1 - \cos{\theta}\right)\frac{\rmd\phi}{\rmd{z}}\rmd{z}.
\end{equation} 
This expression is the same in spherical coordinates as its would be in Euler angles. The the third rotation $\psi$ would be assigned to rate at which the vector potential rotates about the axis (see \sref{poldiscussion}).
 
We can interpret this geometrically. Consider an observer in the moving reference frame of the monopole as it moves up the field line (up in terms of $\hat{z}$). This observer would note a continuous change in the shape of the potential field generated by the flux point source (assuming $\axiscurve$ is not planar). The change in the potential field is due entirely to the geometry of the polymer's axis. It is the change in this potential which shall be used to evaluate the writhing of this curve section. A means by which this observer could track the the fields changing geometry would be to track a point whose relationship to the observer is fixed, in this case the tangent vector. The expression \eref{northspherical} represents the rate of continuous change of the field as the magnetic monopole is transported along the axis $\axiscurve$. We integrate  over the range $z\in[z_{min},z_{max}]$  in order to obtain the total change in the axial geometry, we divide this by $2\pi$ to obtain a measure of the writhing of the section $\axiscurve$
\begin{equation}
\writhe_{pl}(\axiscurve,z_{min},z_{max}) =\frac{1}{2\pi}\int_{z_{min}}^{z_{max}} \left(1 - \cos{\theta}\right)\frac{\rmd\phi}{\rmd{z}}\rmd{z}.
\end{equation} 
This is exactly the directional writhing expression used by both Fain \textit{et al} \cite{fain} and Bouchiat and M\,ezard \cite{bandm} to model the axial geometry of the constrained DNA molecule. In particular Bouchiat and M\'ezard note the issue of the singularity and define their partition function model such that the direction $-\hat{z}$ is avoided using a cut off potential.

We take a different route to handling the singularity, we choose to define an alternative potential for sections of curve with a negative $z$ gradient. We consider a section of curve $\axiscurve$, again representing the polymers axis and whose gradient is negative along $\hat{z}$. In this case we begin the monopole at the maximum $z$ value in the range in which this curve section is defined. The vector potential we attach to our monopole will be 
\begin{equation}
\vec{A}^S = -\frac{\hat{z}\cross\vec{a}}{\vert \vec{a}\vert\,(\vert\vec{a}\vert-\hat{z}\cdot \vec{a})}.
\end{equation}
Again set $\vec{a} =\tvec$ and take the dot product with an infinitesimal change in the tangent direction
\begin{equation}
\label{fieldintsouth} 
\vec{A}^S\cdot\rmd{\tvec} =-\frac{\hat{z}\cross\tvec}{(1-\hat{z}\cdot \tvec)}\cdot \rmd{\tvec}.
\end{equation}
In spherical coordinates we have
\begin{equation}
\vec{A}^S\cdot\rmd{\tvec} = -\left(1+\cos{\theta}\right)\frac{\rmd\phi}{\rmd z}\rmd{z}.
\end{equation}
We integrate this over the range 
$z\in[z_{max},z_{min}]$ in order to recover the directional writhing of this section.
\begin{equation}
\label{localpolar}
\writhe_{pl}(\axiscurve,z_{min},z_{max}) = \frac{1}{2\pi} \int_{z_{max}}^{z_{min}}\vec{A}^S\cdot\rmd{\tvec} = \frac{1}{2\pi} \int_{z_{min}}^{z_{max}}\left(1+\cos{\theta}\right)\frac{\rmd\phi}{\rmd z}\rmd{z}.
\end{equation}
Where we have reversed the integral's limits on the far right of \eref{localpolar}.
\subsection{The non local contribution to the writhing}
It was shown by Wu and Yang 1975 \cite{wuyang} that we can avoid the singular representations by ascribing the potential
$\vec{A}^N$ for sections of $\axiscurve$ with positive gradients along $\hat{z}$ and $\vec{A}^S$ for sections whose gradient along $\hat{z}$ are negative. In order to make this system consistent they show it is necessary to apply a gauge transformation in regions of overlap. In this case we must consider regions which share mutual $z$ ranges, we must define a gauge transformation for each $z$ value, that is $\theta=\pi/2$. For such regions we employ the gauge transformation (in spherical coordinates),
\begin{equation}
\vec{A}^N- \vec{A}^S = \grad A^{eq}  = 2\frac{\rmd\phi}{\rmd{z}}\rmd{z},
\end{equation}
What meaning could be ascribed to a line integral the gauge transformation  $\grad A^{eq}$  for a range of overlap $z\in[z_{min},z_{max}]$? Consider a general curve which has sections moving both positively and negatively about $\hat{z}$. This will naturally mean there are sections of curve which wind about each other about the $\hat{z}$ axis. Consider two sections of $\axiscurve$, $\axiscurve_i$ and $\axiscurve_j$ which share this mutual range $z\in[z_1,z_2]$. We define a vector  $\vec{r_{ij}}(z) = \axiscurve_j(z) - \axiscurve_i(z)$ and denote its orientation in the $x$-$y$ plane as $\Theta(z)$ (the orientation from the $\hat{x}$ axis). This is of course the vector used to define the net winding \eref{netwinding}, excepting that the two curve sections are part of the same curve $\axiscurve$, indeed we will be defining the self winding of these two sections. The rate of rotation of $\vec{r}$ is
\begin{equation}
 \grad\vec{A}^{eq}\cdot\vec{r} =  \frac{\hat{z}\cdot\vec{r_{ij}}(z)\cross\vec{r_{ij}}'(z)}{\mid\vec r(z)\mid^2}\rmd{z}
\end{equation}
thus, in terms of Euler angles, we have
\begin{equation}
\int_{z_1}^{z_2}\frac{\hat{z}\cdot\vec{r_{ij}}(z)\cross\vec{r_{ij}}'(z)}{\mid\vec{r}_{ij}(z)\mid^2}  \equiv \int_{z_1}^{z_2}\left(\frac{\rmd{\psi_{ij}}}{\rmd{z}} + \frac{\rmd\phi_{ij}}{\rmd{z}}\right)\rmd{z}. 
\end{equation}
As with the net winding we consider the contribution due to $\vec{r}_{ji}$ which will be identical. This integral gives the net angle through which the vector potential $\hat{z}\cross\vec{r}_{ij}$ winds, about the $\hat{z}$ axis. We call this term the non local polar writhing $\writhe_{pnl}$ as is defined in terms of a vector potential linking distinct sections of the curve $\axiscurve$. 

\subsection{The Polar writhe}\label{poldiscussion}
Consider the axis of a ribbon, represented by the curve $\axiscurve$, which spans a range $z\in[z_{min},z_{max}]$ and  which has $n-1$ turning points along $\hat{z}$  ($\frac{\rmd\axiscurve}{\rmd z} = 0$), and  which can be split into $n$ sections. We can define the net local polar writhing as
\begin{equation}
\writhe_{pl}(\axiscurve,z_{min},z_{max}) = \frac{1}{2\pi}\int_{z_i^{min}}^{z_i^{max}}(1-\vert\cos{\theta_i}\vert)\frac{\rmd\phi_i}{\rmd{z}}\rmd{z}.
\end{equation} 
We further define the net non-local polar writhing,  $\writhe_{pnl}$. Consider sections of the curve $\axiscurve_i$ and $\axiscurve_j$, which share a mutual range of $z$ values $z\in[z_{ij}^{min},z_{ij}^{max}]$, we have
\begin{equation}
\writhe_{pnl}(\axiscurve,z_{min},z_{max}) = \mutualsum injn \frac{\sigma_i\sigma_j}{2\pi}\int_{z_{ij}^{min}}^{z_{ij}^{max}}\left(\frac{\rmd{\psi_{ij}}}{\rmd{z}} + \frac{\rmd\phi_{ij}}{\rmd{z}}\right)\rmd{z}.
\end{equation}
Finally we define the polar writhing $\writhe_p$, of the curve, as the sum of these two contributions 
\begin{equation}
\writhe_p(\axiscurve,z_{min},z_{max})  = \writhe_{pl}(\axiscurve,z_{min},z_{max})  + \writhe_{pnl}(\axiscurve,z_{min},z_{max}).
\end{equation}
\subsection{A comparison the the writhe result of Bouchiat and M\'ezard and Fain \textit{ et al}}
We can compare this result to the expression used by Bouchiat and M\'ezard in order to understand the various components of the polar writhe measure. Consider first a configuration whose gradient along $\hat{z}$ is always positive. In this case we would only need to apply the vector potential $\vec{A}^N$ and the polar writhe is defined entirely using the vector potential $\vec{A}^N$. In effect the tip of $\vec{A}^N$ would draw out an imaginary curve $\secondary$ creating a ribbon $R(\axiscurve,\secondary)$. Such a ribbon would have $\lwig$ and twisting measures. We can define the local twisting as the integral over the rate $\omega_1$ \eref{omega1polar} divided by $2\pi$ 
 \begin{equation}
\twist(R,z_{min},z_{max}) = \frac{1}{2\pi}\int_{z_{min}}^{z_{max}}\left(\frac{\rmd{\psi}}{\rmd z}\  + \vert\cos{\theta}\vert\,\frac{\rmd{\phi}}{\rmd{z}}\right)\rmd{z}.
\end{equation}
We note the difference between the net winding (here given entirely by its local contributions) and the twisting gives the required polar writhing expression in this case, that is 
\begin{eqnarray}
\label{localcalug}
\lwig_l(R,z_{min},z_{max}) - \twist(R,z_{min},z_{max}) = \writhe_{pl}(R,z_{min},z_{max}).
\end{eqnarray}
This result was also obtained by Bouchiat and M\'ezard. However, differences arise for polymer configurations for which there are downward travelling components. We note that, due to the $\vert\cos{\theta}\vert$ in our twisting expression \eref{localcalug} is also true of the $\vec{A}^S$ contributions to $\writhe_{pl}$. This is the first major difference between this directional writhing formulation in contrast with the work of Bouchiat and M\'ezard. In this case northern and southern local writhing are given equal weighting.

The second difference results from the contribution to the writhing resulting from the gauge transformation $\grad\vec{A}^{eq}$. A polymer with at least one turning point along $\hat{z}$  will have a non local writhing contribution. Rossetto and Maggs \cite{rossetto} and Neukirch and Starostin \cite{nands} have already noted that curves with non local writhings will be inappropriately evaluated by the writhing expression \eref{fuller2z} proposed by Bouchiat and M\'ezard, though they make a comparison with the $\writhe$ as measured by \eref{writhe}. Rossetto and Maggs argue that the Bouchiat and M\'ezard model works well for cases in which the vertical force $F$ is sufficiently high, such that geometrical configurations with non local writhing (or winding) lead to a statistically negligible contribution. We see in \sref{writhecomp} that the Bouchiat and M\'ezard writhing expression actually  evaluates a certain proportion of polymer configurations which have non-local winding appropriately, that is it returns the same evaluation as the polar writhe.

\subsection{A comparison of the various writhing expressions}\label{writhecomp}
We have at several points in this note encountered various writhing expressions. We have the double integral writhing expression defined by equation \eref{writhe} which we shall denote $\writhe$, the directional writhing expression used by Bouchiat and M\'ezard, here denoted $\writhe_z$, in their model, finally, we have the polar writhe expression introduced above. We use to example curve studies in order to demonstrate that the properties of the polar writhe measure are those one would expect of the possible geometrical configurations obtainable by a constrained elastic polymer. The procedure will involve tracking the evolution of $\writhe$, $\writhe_z$ and $\writhe_p$, as applied to a set of curves over a period $[0,t]$, as $t$ is increased.

\subsection{Case 1 - Helical configurations}
 \centrefig{10cm}{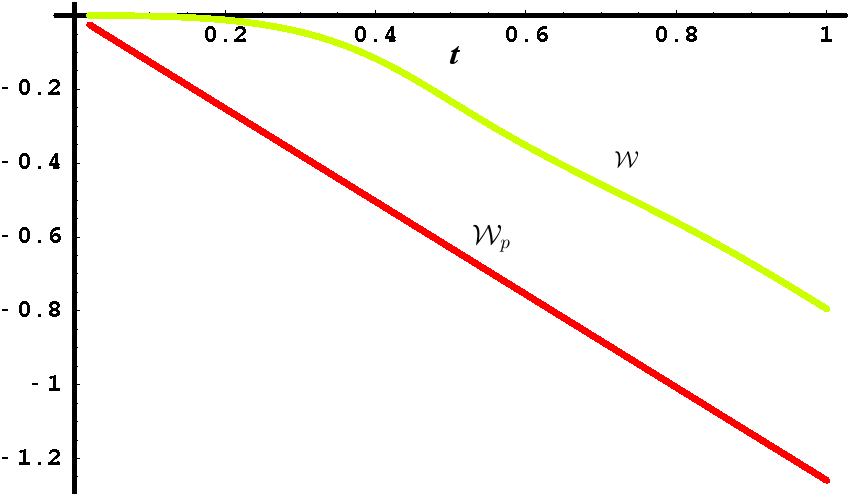}{\label{helixfulgauss}A plot of $\writhe(\axiscurve)$ and $\writhe_p[\axiscurve(t)]$, where $\axiscurve = (\sin{4\pi t},\cos{4\pi t},t)$, evaluated over $[0,t]$ for $t\in[0,1]$. The $\writhe_p$ plot is linear, the $\writhe$ becomes linear after $t\approx1.4$.}

Helical shapes are a common configuration assumed by symmetric elastic rod or polymer, under the force torque pair. Consider a rectilinear elastic rod which central axis is place on $\hat{z}$, a scenario similar to the constrained DNA experiments we have discussed. It is known that a combination of deformations including a force along the axis and a torque perpendicular to $\hat{z}$ will causes buckling modes. One of the simpler configurations assume would be a helical shape, that is to say the curves axis is helical (see for example Healey and Mehta \cite{handm}). Indeed, using the same partition function model as Bouchiat and M\'ezard, Fain \textit{et al} \cite{fain} demonstrated that some of the possible energy configurations of the constrained DNA polymer we have described can be helical. So the appropriate evaluation of the writhing of helical polymer configurations is a desirable properties of our measure.  On might imagine that the greater the applied torque the greater the possible number of turns of the helix. We would expect the writhing to increase linearly with the number of turns for a symmetric helix. 

Figure \ref{helixfulgauss} depicts the result of evaluating $\writhe$ and $\writhe_p$ of a helix ($\axiscurve(t) = (\sin{4\pi t},\cos{4\pi t},t)$) over the period $[0,1]$  for $t$ values $t\in[0,1]$. This helix winds about a fixed direction ($\hat{z}$) and could for example represent the axis of a supercoiled DNA molecule bound between two surfaces. We see a marked difference in the writhing interpretation of each measure (note that the results for $\writhe_p$ in this case would be the same as $\writhe_z$ as the helical always has a positive vertical gradient). Specifically $\writhe_p$ increases linearly with $t$. $\writhe$ increases slower at first until it starts to increase linearly after roughly $t=0.4$. The reason for this difference derives from their interpretation of non-local writhing. $\writhe_p(\axiscurve)$ in this scenario records no non-local writhing, over the full parametrization range. $\writhe(\axiscurve)$ however records non-local windings from all possible view points, thus will record the helix as exhibiting non-local windings. So we can see that both $\writhe_z$ and $\writhe_p$ give the appropriate measure in such cases, $\writhe$ however does not. To reiterate this is because the helix is a curve whose contortion is directionally specific (about $\hat{z}$) in this case. The non local writhe $\writhe$ is ill equipped to evaluative such curves as much of the directional specificity of its geometry will be lost in the averaging process.

\begin{figure}
\begin{center}
\includegraphics[width=4cm]{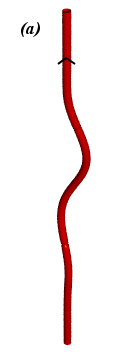}
\includegraphics[width=4cm]{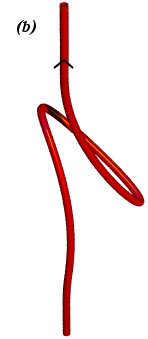}
\includegraphics[width=4cm]{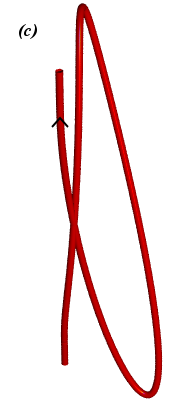}
\end{center}
\caption[Various depictions of a curve (given by (\ref{polcv})) which begins as a straight line and kinks about its centre]{\label{polymer} The curve $polcv(t)$, equation(\ref{polcv}), evaluated over a period $t\in[-1,1]$, for values of $a=0.2$ (a), $a = 0.8$ (b) and $a= 1.16$ (c) respectively (note the $z$ component has been multiplied by a factor of $10$ for the sake of aesthetics). In (a) the curve has developed a small sigmoidal kink about its midriff. This kink develops such that the curve develops $\writhe_{pnl}$ contributions about its middle section. In (c) the curves actually intersect, forming a double point. For $a>1.1.6$ the curve will have passed through itself leading to a change in its non-local writhing of $+2$ (Figure \ref{polymerAll}).}
\end{figure}
\subsection{Case 2 - non locally winding configurations}
We now define a curve (depicted in Figure \ref{polymer}), which begins as a curve winding locally along $\hat{z}$. The 
coiled
 section then drops back over its axis leading to non-local winding of the curve about itself. It can be parameterized using the following set (we call the set polcv, short for polymer-type curve)
\begin{eqnarray}\label{polcv}
\label{error}erf(t) = \frac{2}{\sqrt{\pi}}\int_0^t e^{-t'^2}\rmd{t'}, \\ 
polcv(t) = (a 7.5(erf(-2t^2) +1)\sin{2\pi t}\nonumber\\
, 2 a(erf(-2t^2)+1)\nonumber\\
,5(t+ a 5erf(-2t^2)+1)(t(t-a)(t+a)) ),
\end{eqnarray}
where $a$ is a constant which controls the contortion of the polymer. Increasing $a$ causes a kink to form about the curves midpoint. Note, its endpoints will remain fixed. The evolution of this curve has several specific properties which we would require our writhe measure to capture. it has a transition from modes which have only local writhing about $\hat{z}$ to those for which distinct sections of the curve are wound about each other. As the torque is causing this winding is in the $x$-$y$  axis we would like our measure to reflect this. Finally in (c) \Fref{polymer} the curve passes through itself. This the kind of deformation which would be forbidden in the DNA experiments. We would like our measure to reflect this fact in some way. 

\centrefig{12cm}{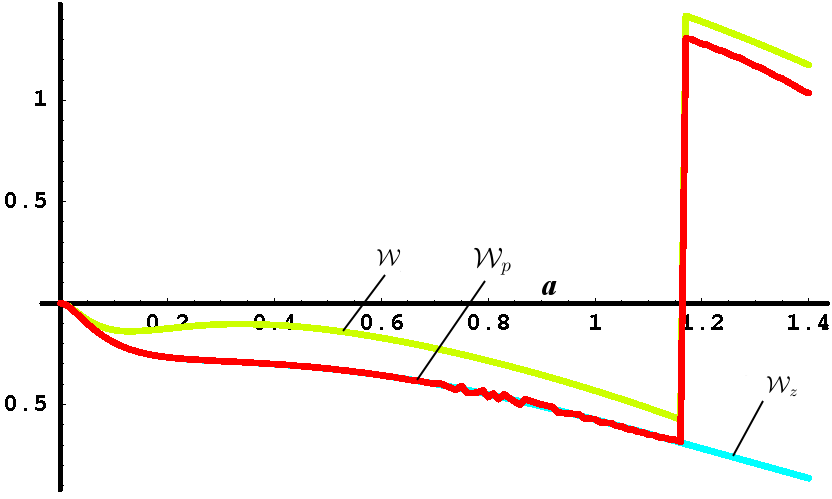}{\label{polymerAll} Plots of $\writhe$, $\writhe_{z}$ and $\writhe_p$ against $t$ for the curve $polcv(t)$, evaluated over a period $t\in[-1,1]$, over a range of a values $a\in[0,1.4]$. For $a\approx[0,1.175]$ $\writhe_p$ and $\writhe_z$ agree in their evaluations. However at $a \approx1.75$ there is a jump in value of both $\writhe$ and $\writhe_p$ as the curve crosses itself. This jump does not occur for $\writhe_z$.}
\centrefig{12cm}{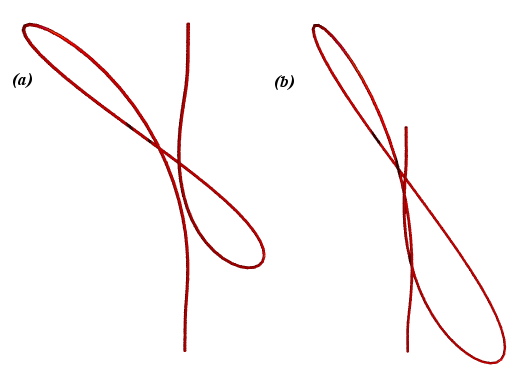}{\label{polcurve} Figure (a) represents $polcv(t)$ (\ref{polcv}), plotted over a range $t\in[-1,1]$, with $a=1.1$. It represents an example of the curve before it has passed through itself (we would expect $\writhe_p$ and $\writhe_z$ to return the same measure, and they do (Figure \ref{polymerAll}). (b) is the same curve with $a =1.2$; the curve has passed through itself.}

Figure \ref{polymerAll} details the results for $\writhe_p[\axiscurve(t)]$, $\writhe_z[\axiscurve(t)]$ and $\writhe[\axiscurve(t)]$, as evaluated over the period $t\in[-1,1]$, for $a\in[0,1.4]$. There is an agreement in the writhing interpretations of $\writhe_{z}$ and $\writhe_p$ up until a value of $a = 1.16$. This includes the period in which non-local windings are formed. This occurs as the non local component $\writhe_{pnl}$ is such that the sum effectively cancels out $\writhe_{pl}$ components for sections of the curve whose gradient along $\hat{z}$ is negative (see Berger and Prior  \cite{main}. This is intriguing as it would offer a possible explanation as to why the models of Bouchiat and M\'ezard (and Moroz and Nelson \cite{mandn}) performed better than one might expect. We note the $\writhe$ expression evaluates these curves differently. Once again this is due to the fact that the $\writhe$ measure averages aver all directions.

After $a=1.16$ there is a jump in the values of $\writhe$ and $\writhe_p$ of $+2$, as a result of the curve passing through itself.  $\writhe_p$ and $\writhe$ detect the crossing by registering a jump of $+2$. $\writhe_{z}$ on the other hand recognizes only a small difference in writhing between two such states. This will be a key factor in the partition function model we posit in \sref{partition}. The discontinuous jump in writhing due to the curve being passed through itself could not be compensated for by a change in twisting, as the twisting changes continuously. Thus this change would lead to a change in linking. 

The example studies in this chapter have demonstrated that the polar writhe expression satisfies the requirements of key geometrical aspects of possible constrained polymer configurations, especially with regards to the DNA micro-manipulation experiments in which we are interested. In particular we have seen it appropriately captures the directional specificity of the forces applied to this system and the self avoidance condition inherent to the molecules. 

\section{Linking and the partition function}\label{partition}
The general form of the partition function for an unconstrained molecule would be.
\begin{equation}
Z = \int \mathcal{D}(\theta,\phi,\psi)\,\,e^{\frac{-E}{k_b T}}, 
\end{equation}
Where $\mathcal{D}(\theta,\phi,\psi)$ is the functional space over all possible paths. However we must place certain restrictions on the allowed configurations taken by the molecule. As discussed in section \sref{lkinvariant} this constraint is the net winding. This can be split into the local linking of sections for which $i=j$ and non local sections where $i\neq j$. This constraint is applied  using the dirac delta function as in \cite{bandm}. 
\begin{equation}
Z_{\lwig} = \int \mathcal{D}(\theta,\phi,\psi)\,\delta\left(\lwig  - \lwig_{l} - \lwig_{nl}\right) \,e^{\frac{-E}{k_b T}}, 
\end{equation}
The partition function has two inputs the applied winding $\lwig$ and the applied stretching force $F$. It was shown in Berger and Prior that the local linking contribution can be split into twisting and writhing components with the writhing component equal to the local polar writhe $\writhe_{pl}$ and the twist given as the integral over the angular rotation $\omega_1$ \eref{omega1polar} (divided by $2\pi$). For a ribbon $R(\axiscurve,\secondary)$, parametrized by $\hat{z}$, with $n$ turning points about $\axiscurve$ we have
 \begin{equation}
\twist(R,z_{min},z_{max}) = \frac{1}{2\pi}\sum_{i=1}^{n}\int_{z_i^{min}}^{z_i^{max}}\sigma_i\left(\frac{\rmd{\psi}}{\rmd z}\  + \vert\cos{\theta}\vert\,\frac{\rmd{\phi}}{\rmd{z}}\right)\rmd{z}.
\end{equation}
So
\begin{equation}
\link_{l}(R,z_{min},z_{max}) = \twist(R,z_{min},z_{max}) + \writhe_{pl}(\axiscurve,z_{min},z_{max})
\end{equation}
We now make a final observation. Consider the energy expression which is an integral over the densities defined in \sref{energydensity} 
\begin{equation}
\label{energy}
E = \sum_{i=1}^{n}\int_{z_i^{min}}^{z_i^{max}}E'_{bend}(z)+E'_{twist}(z) +  E'_{stretch}(z)\,\rmd{z}
\end{equation}
All components of this expression are local. The non local winding only contributes to the energy by limiting the set of allowed configurations the polymer can exhibit. Thus we can separate the integral, performing a local partition function for a fixed amount of non-local winding ($\lwig_{nl}$),
\begin{equation}
\label{localpartition}
Z(\lwig,F,\lwig_{nl}) = \int \mathcal{D}(\theta,\phi,\psi)\delta\left((\lwig - \lwig_{nl}) - \writhe_{pl} - \twist\right)e^{\frac{E}{k_b T}}
\end{equation}
One can then integrate over the range of $\lwig_{nl}$ values (note this is theoretically unbound) to give us the following.
\begin{equation}
\label{nonlocalpartition}
Z(\lwig,F) = \int_{-\infty}^{\infty}Z(\lwig,F,\lwig_{nl})\rmd{\lwig_{nl}}.
\end{equation}
The pair of equations \eref{localpartition} and \eref{nonlocalpartition} define a self avoiding partition function. 
\section{Conclusions}
We have discussed the geometrical modelling of physical structures which can be represented mathematically as ribbon's or rod's, and which are constrained at their endpoints. Further they are subject to the condition that all deformations are restricted such that they are contained between two planes at the curves endpoints. In particular the following results were detailed.
\begin{itemize}
\item We defined the correct components of the ribbon's energy expression, in terms of an orthonormal framing of the axis of the ribbon. This differs form previous work (\cite{fain},\cite{bandm},\cite{mandn}) in that it correctly defined the rates for sections moving either upwards (along $\hat{z}$) and downwards. The downwards moving sections had previously been evaluated incorrectly. Despite this the energy expression was shown to be the same. See \sref{energydensity}.
\item We demonstrated that the net winding was the correct topological constraint required to restrict the allowed configurations of the ribbon. This expression can be defined as as decomposition of local and non-local components. See \sref{lkinvariant}.
\item We derived the correct directional writhing expression for constrained ribbons.   This derivation used the correct form of a transported magnetic monopole (or unit of magnetic flux), this expression is termed the polar writhe (it has been derived before in a different fashion \cite{main}). This derivation neatly links the problem of a ribbon, manipulated by directionally inclined torsional and stretching forces, to the quantum problem of a top in an external magnetic field. This builds upon the insight of Bouchiat and M\'ezard, who first suggested this link, by detailing the correct formulation of the monopole filed. Using example curve studies we demonstrated that the polar writhe has the required geometrical properties for the modelling of constrained ribbons. See \sref{polarmonopole}.
\item Using the above results we derive a partition function model for the constrained ribbon which forces the constraint of self avoidance using the net winding measure. Further, using the decomposition of the net winding into local and non-local components, we modified this model such that it allows for the possibility of an analytic treatment, see \sref{partition}.  
\end{itemize}  
This work represents an improvement to the previous work, discussed in the introduction, in that it includes the necessary physical constraint of self-avoidance. The next step would be to approach this model using analytical and numerical techniques. This could involve minimization of the energy expression \eref{energy}, subject to the net winding constraint, in a similar manner to that detailed by Fain et al \cite{fain}. Once could also follow the leads of Bouchiat and M\'ezard \cite{bandm} and Moroz and Nelson \cite{mandn} in tackling the partition function.

\appendix
\section{Problems with the symmetric top model and twisting}\label{symtop}
\centrefig{6cm}{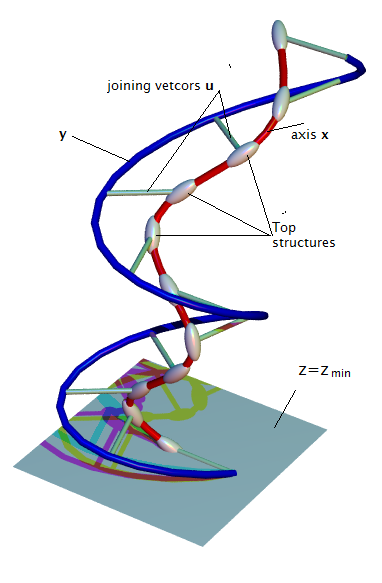}{\label{toptwist}Depicted is the process by which an infinitesimal top with an attached vector can be used to draw out a polymer consisting of an axis $\axiscurve$ and a surrounding curve $\secondary$ which twists about the axis. This set up could represent a constrained DNA molecule. The top begins oriented along the tangent direction of $\axiscurve$ in the plane $z = z_{min}$, it is then moved up the axis with the necessary Euler rotations applied in order for it to be aligned along the tangent direction of $\axiscurve$. In addition Euler rotations are applied such that the joining vector $\vec{u}$ stays attached to $\secondary$. This set of Euler rotations can be used to evaluate the twisting of $\secondary$ about $\axiscurve$.}
A key feature of the work of Bouchiat and M\'ezard is the mapping of the polymer chain problem to the quantum mechanical model of a symmetric top, subjected to a magnetic field, through an imaginary time transformation (effectively a reversal of the Wick rotation). If we assume the molecule is symmetric about its axis then the molecule can be modelled as a ribbon which consists of a curve $\axiscurve$, representing the axis of the molecule and a surrounding curve $\secondary$ which can be used to represent one of the phosphate back bones of the DNA molecule (see \sref{toptwist}). In order to define the geometry  of this molecule we must consider both the contortion (bending and winding) of $\axiscurve$, in other words the writhing, and the extent to which $\secondary$ wraps around $\axiscurve$, the twisting. 

\centrefig{5cm}{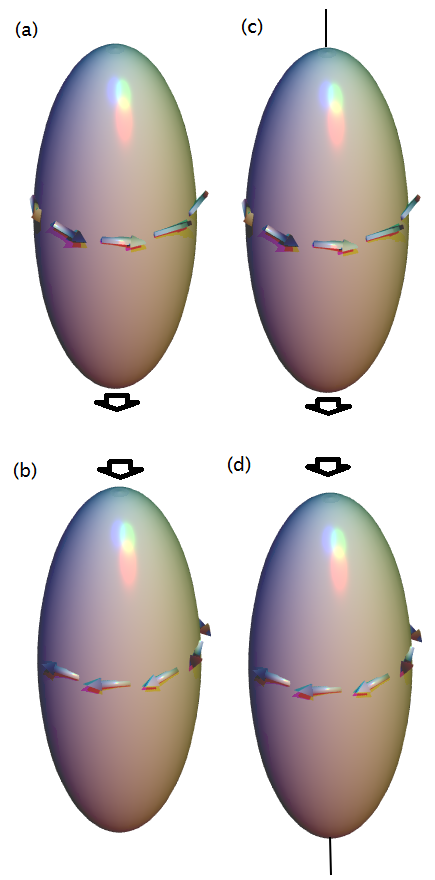}{\label{topsymmetry}A depiction of the failure of the symmetric top to distinguish the appropriate handedness of twisting. Figure (a) is a symmetric top it has had a rotation applied to it in the x-y plane such that it is spinning in the direction shown. The top depicted in (b) represents the top in (a) rotated through $\pi$ radians about the $x$ axis. To the observer it would appear that (b) has the opposite rotation to that of (a). Figure (c) is a top with the same rotation applied to it as (a), as shown by the arrows depicting its rotation. However, (c) is not symmetric in that its top point has a marker attached to it, making it distinct. We rotate (c) through $\pi$ radians (the same rotation as (a) to (b)) in order to produce (d). We can now identify that (c) and (d) have the same sense of rotation (right handed about the direction point by our marker).}

We can link this structure and the a top using Euler angle rotations to see why the quantum analogy used by Bouchiat and M\'ezard had some success. Let us consider a symmetric top (see (a) in \fref{topsymmetry}), consider it shrunk to an infinitesimal size with a vector $\vec{v}$  attached to the molecule and lying in the $x-y$ plane, this vector is fixed and will rotate and with the top. This construction is placed at the lowest point of the molecule (\fref{toptwist}) such that the tip of $\vec{u}$ lies on $\secondary$.  We now move the top along the molecule applying the necessary Euler rotations such that its tip points along the tangent direction of $\axiscurve$ and $\vec{u}$ will be rotated such that it remains on $\secondary$ (\fref{toptwist}). Using this set up we can define the geometry of the curve in terms of the unique set of Euler angle rotations applied to the top. There is, however, a problem with this set-up.

The issue relates to the twisting of the molecule and the top. The twisting of the molecule can be quantified by tracking the rotations, in the $x$-$y$ plane, of the vector $\vec{u}$. The key issue is with the use of the Euler rotation $\theta$. The symmetric top is by definition symmetric about its centre line, which in our set up always lies along the tangent direction of $\axiscurve$. Imagine spinning the top about its axis when its tip is pointing directly upwards and spinning it about its axis in a right handed manner (i.e. it follows the right hand thumb rule) (see  (a) in \fref{topsymmetry}). We ask an observer to note the orientation with which it spins. We then (unknown to the observer) turn the top such that the tip which was pointing upwards now points downwards (in Euler rotations this would require a rotation $\theta = \pi$). To the observer it would appear that the top is spinning in the opposite direction (see (b) \fref{topsymmetry}). In the case of the top representing the polymer molecule this would be a problem. No new rotation about $\hat{z}$ has been applied so the twisting (rotation of the top about its axis of symmetry) of the molecule should be the same. However in the symmetric top case the twisting appears to have reversed. We need a top which is in some way marked (as in (c) \fref{topsymmetry}) such, such that we can tell the top from the bottom. In this case the observer would be able to note that the top described above has the same spin direction after the polar rotation has been applied (compare (c) to (d) in \fref{topsymmetry}). In terms of Euler angle rotation this is achieved by using $\vert\cos{\theta}\vert$ rather than $\cos{\theta}$ in our calculations. The modulus sign will mean that there is a symmetry in the twisting of the top about $\theta = \pi/2$.


\section*{Bibliography}
\begin{thebibliography}{99}
\bibitem{aldinger} Aldinger J, Klapper I, Tabor M, 1995, \emph{J. Knot. Theory. Ram, }\boldmath{4(3)}, 343.
\bibitem{dnabook} Bates A.D, Maxwell A, 2005, \emph{Oxford University Press}.
\bibitem{bergerfield} Berger M A, Field G B, 1984, \jfm, \boldmath{147}, 133.
\bibitem{main} Berger M A, Prior C B, 2006, \jpa, \boldmath{39}, 8321.
\bibitem{bandm} Bouchiat C, M\'ezard M, 2000, \emph{The European Physical Journal E,} \boldmath{4(2)}, 377.
\bibitem{boumezlett} Bouchiat C, M\'ezard, 2002, \prl, \boldmath{88(8)}, 089802.
\bibitem{brereton1} Brereton M G, Shah S, 1982, \jpa, \boldmath{15(3)}, 985.
\bibitem{bmexp2} Bustamante C, Marko F D, Sigga E D, Smith S, 1994, \emph{Science, } \boldmath{265}, 1599.
\bibitem{dnareview} Bustamante C, Bryant Z, Smith S B, 2003, \emph{Nature, } \boldmath{421}, 423.
\bibitem{calgal1} \calug G, 1959, \emph{Czechoslovak. Math. J, }\boldmath{66}, 588.
\bibitem{calgal2} \calug G, 1961, \emph{Comm. Acad R.P. Romine, } \boldmath{11}, 829.
\bibitem{dirac} Dirac P M, 1931, \emph{Proc.R.Soc, }\boldmath{133} 60.
\bibitem{epple} Epple M, 1998, \emph{Mathematical Intelligencer,} \boldmath{3}, 20.
\bibitem{fain} Fain B, Rudnick J, \"{O}stlund S, 1997, \emph{Phys. Rev. E, } \boldmath{55(6)}, 7364.
\bibitem{fuller0} Fuller F B, 1971, \emph{Proc. Nat. Acad. Sci USA, } \boldmath{68(4)}, 815.
\bibitem{fuller} Fuller F B, 1978, \emph{Proc. Natl. Acad. Sci. USA, } \boldmath{75(8)}, 3557.
\bibitem{handm} Healey T J, Mehta P G, 2004, \emph{Jorunal of Bifurcation and Chaos}, \emph{15(3)}, 949.
\bibitem{hirsch} Hirsch M W, 1976, \emph{Pub: Springer Verlag, New York}.
\bibitem{mandv}  Vologodskii A V, Marko J F, 1997, \emph{Biophysics J., } \boldmath{73(1)}, 123.
\bibitem{mandn} Moroz J, Nelson P, 1998, \emph{Macromolecules, } \boldmath{31}, 6333.
\bibitem{nands} Neukirch S, Starostin E, 2008, \emph{Phys. Rev. E, }\boldmath{78(4)} 041912 .
\bibitem{nands2} Neukirch S, Starostin E, 2009, \emph{Phys. Rev. E} \boldmath{80} 063902.   
\bibitem{pohloriginal} Pohl W F, 1968, \emph{J. Math. and Mech., } \boldmath{17}, 975.
\bibitem{rosmaglett} Rossetto V, Maggs A C, 2002, \prl,\boldmath{88(8)}. 089801.
\bibitem{rossetto} Rosetto V, Maggs A C, 2003, \JCP, \boldmath{118(2)} 9864. 
\bibitem{sinha} Sinha S, 2004, \emph{Phys Rev E, }\boldmath{70(1)} 011801.
\bibitem{indians} Samuel J, Sinha S, Ghosh A, 2006, \emph{J Phys: cond. Matt., } \boldmath{18(14)}, 253.
\bibitem{indians2} Samuel J, Sinha S, Ghosh A, 2009, \emph{arXiv:0905.0250}.
\bibitem{solomon} Solomon B, 1996, \emph{Amer. Math. Monthly, } \boldmath{103},  30.
\bibitem{starostin} Starostin E L, 2005, Chapter \boldmath{26} in \emph{Physical and numerical models in knot theory including applications to the life sciences, } 525.
\bibitem{bmexp1} Smith S B , Finzi L, Bustamante C, 1992, \emph{Science, } \boldmath{258} 1122.
\bibitem{supercoilpic} Travers A, Muskhelishvili G, 2005, \emph{Nature Rev. Microbiology, }\boldmath{3}, 157.
\bibitem{vanderheijden2} van der Heijden G H M, Peletier M A, Planque R, 2007, \emph{Quarterly of Applied Mathematics, } \boldmath{65}, 385.
\bibitem{volosolo} Vologodskii A, 1994, \emph{Macromolecules, }\boldmath{27}, 5623.
\bibitem{white} White J H, 1969, \emph{American Journal of  Mathematics,} \boldmath{91(3)}, 693.
\bibitem{wuyang} Wu T, Yang C, 1975, \emph{Phys.Rev.D}, \boldmath{12(12)} 3845.
\end{thebibliography}
\end{document}